\documentclass[useAMS,usenatbib]{mnras}
\usepackage{graphicx,psfig,cite,epsf}  
\usepackage{times}
\usepackage{subfigure}
\usepackage{textcomp}
\usepackage{multirow}
\usepackage{amsmath}	
\usepackage{amssymb}	
\usepackage{gensymb}

\def\apjl{ApJ}                
\def\apjs{ApJS}               
\def\ao{Appl.~Opt.}           
\def\aap{A\&A}                

\def\src {GRB 160623A}
\def\xmm{{\it XMM-Newton}}

\title[The spectacular case of GRB 160623A]{Behind the dust curtain: the spectacular case of GRB 160623A}

\author[Pintore F. et al.]{F. Pintore$^1$\thanks{E-mail: pintore@iasf-milano.inaf.it}, A. Tiengo$^{1,2,3}$, S. Mereghetti$^1$, G. Vianello$^4$,  R. Salvaterra$^1$,  P. Esposito$^5$, \newauthor E. Costantini$^6$, A. Giuliani$^1$, Z. Bosnjak$^7$ \\
 $^1$ INAF -- IASF Milano, Via E. Bassini 15, 20133 Milano, Italy\\
 $^2$ Scuola Universitaria Superiore IUSS Pavia, Piazza della Vittoria 15, 27100 Pavia, Italy\\
$^3$ Istituto Nazionale di Fisica Nucleare, Sezione di Pavia, Via A. Bassi 6, I-27100 Pavia, Italy \\
 $^4$  SLAC National Accelerator Laboratory, Stanford University, Stanford, CA 94305, USA \\
$^5$ Anton Pannekoek Institute for Astronomy, University of Amsterdam, Postbus 94249, 1090-GE Amsterdam, The Netherlands \\
 $^6$  SRON Netherlands Institute for Space Research, Sorbonnelaan, 2, 3584-CA, Utrecht, The Netherlands   \\
$^7$ Faculty of Electrical Engineering and Computing, University of Zagreb, 10000 Zagreb, Croatia   \\
 }

\date{Accepted  . Received  ;}

\pubyear{2017}

\begin{document}

\maketitle

\begin{abstract}
We report on the X-ray dust-scattering features  observed around the afterglow of the gamma ray burst \src . With an  \xmm\ observation carried out $\sim2$ days after the burst,  we found evidence of at least six  rings, with   angular size expanding between $\sim$2 and 9 arcmin, as expected for X-ray scattering of the prompt GRB  emission by  dust clouds in our Galaxy. 
From the expansion rate of the rings, we measured the distances of the dust layers with extraordinary precision: $528.1\pm 1.2$ pc, $679.2 \pm 1.9$ pc, $789.0 \pm 2.8$ pc, $952 \pm 5$ pc, $1539 \pm 20$ pc and $5079 \pm 64$ pc.  A spectral  analysis of the ring spectra, based on an appropriate dust-scattering model (BARE-GR-B from \citealt{zubko04}) and the estimated burst fluence, allowed us to derive the column density of the individual dust layers, which are in the range $7\times10^{20}-1.5\times10^{22}$ cm$^{-2}$. The farthest dust-layer (i.e. the one responsible for the smallest ring) is also the one with the lowest column density and it is possibly very extended, indicating a diffuse dust region.
The properties derived for the six dust-layers (distance, thickness, and optical depth) are generally in good agreement with independent information on the reddening along this line of sight and on the distribution of molecular and atomic gas.

\end{abstract}

\section{Introduction}

Since the advent of      X-ray observatories with good imaging capabilities and high throughput, such as \xmm, {\it Chandra} and {\it Swift}, an increasing number of dust-scattering halos around bright X-ray sources has been studied in detail. The physical process responsible for the halos is the small-angle scattering of soft X-ray photons on dust grains in the interstellar medium. 
Scattered photons form a diffuse halo around the central bright source in the revealed images and, due to their longer path length  (which increases with the scattering angle and depends on the distances of the scattering dust from the X-ray source and the observer), are detected with a time delay with respect to the unscattered photons. Therefore, changes in the intensity and radial profile of the halo can be seen in the case of  variable  sources. 
A particularly simple case is provided by impulsive sources which are very bright only for a short time interval, such as gamma-ray bursts (GRBs), bursts from magnetars and type I X-ray bursts from accreting neutron stars. 
The halo characteristics are also affected by the distribution of the dust along the line of sight. 
For narrow dust clouds, the halos produced by impulsive sources appear as  rings.
 
Halos and expanding rings have been  detected around Galactic binary systems \citep[e.g.][]{heinz15, heinz16, vasilopoulos16}, magnetars \citep[e.g.][]{tiengo10, svirski11,pintore17} and GRBs \citep[e.g.][]{vaughan04,tiengo06,vianello07}.  
The study of the  energy- and time-dependence of  scattering halos can provide important information on the chemical composition, grain size, and spatial distribution of the dust, as well as on the distance of the X-ray sources \citep[e.g.][]{trumper73,mathis91,mir99,predehl00,draine03,costantini05}. 

In the case of X-ray photons of  a burst source at distance  {\it d} scattering on a narrow layer of dust   at distance  {\it d$_{dust}$}, the ring angular radius $\theta(t)$ can be expressed as:
\begin{equation}
\theta(t) = \bigg[ \cfrac{2c}{d} \cfrac{(1-x)}{x} (t-\text{T}_0) \bigg]^{0.5},
\end{equation}
where {\it x} = $d_{dust}/d$, {\it c} is the speed of light and T$_0$ is the time of the burst. When  $d \gg d_{dust}$, as for  GRBs scattering on  dust clouds in our Galaxy, the above expression simplifies to:
\begin{equation}
\theta(t) = \bigg[ \cfrac{2c (t-\text{T}_0)}{d_{dust}} \bigg]^{0.5} = K (t-\text{T}_0)^{0.5},
\end{equation}
where the degeneracy between the source and the dust-layer distances is removed.
\noindent
Therefore, by measuring the rate of the angular expansion of the ring, one can obtain a very accurate, model-independent,   measure of the dust-layer distance.
The scattering rings may also have an intrinsic width  $\Delta \theta$, which depends on the thickness of the dust-layer and on the duration of the impulse emission.  The observed width is also affected by the finite angular resolution of the instrument and by the ring expansion during the integration time needed to detect it.

In this work, we study the bright dust-scattering rings produced by  \src . This GRB, discovered by the Fermi/GBM instrument and detected also above 1 GeV with the Fermi/LAT \citep{vianello16}, was located at Galactic coordinates $l$ =  84.2$^{\circ}$, $b$ = --2.7$^{\circ}$ that is a direction crossing a long path of our Galaxy, where dust clouds are expected. 
\src\  was detected also by Konus-Wind, with a start time at  T$_0$ = 04:59:37.594 UT.
The  50 keV -- 15 MeV spectrum of the burst,  averaged  from T$_0$ to T$_0+38.912$ s, was well fit  by a {\it Band} function with $\alpha$ = $-1.05\pm 0.03$, $\beta$ = $-2.67\pm 0.1$, and peak energy  E$_p$ = $562\pm23$ keV; the   fluence in the 20 keV--10 MeV range was  $(6.6\pm0.1)\times10^{-4}$ erg cm$^{-2}$ \citep{frederiks16}.  {\it Swift}/XRT observations of  \src\ started $\sim40$ ks after T$_0$,  and  showed an  X-ray afterglow surrounded by a dust-scattering ring   with  radius of $\sim$3.5 arcmin \citep{tiengo16}.  A quick look analysis of the XRT data indicated that the ring was produced by at least one Galactic dust cloud   at a distance of $\sim$800 pc.

The optical afterglow of \src\ was observed by the Nordic Optical Telescope (NOT) and the Gran Telescopio Canarias (GTC) \citep{malesani16,Castrotirado16}. H$\alpha$, S II and N II emission lines were detected at a redshift of z $=$ 0.367, corresponding to a distance of $\sim$1530 Mpc. 

After finding the evidence of dust scattering rings in the {\it Swift}/XRT data, we triggered an 
\xmm\ target of opportunity observation to study the rings in more detail and characterize the properties of the dust clouds in this direction.

\section{data reduction}
\label{data_reduction}

The {\it XMM-Newton} observation of  \src\ started  at 21:04 UT of 2016 June 24,   $\sim2$ days after the burst, and lasted about 16 hours.
We used the data obtained with the  pn \citep{struder01short} and the two MOS  cameras \citep{turner01short} of the EPIC instrument. They were operated in full-frame mode, providing 0.2--12 keV images over a field of view of $\sim27'\times27'$,  and with a thin optical blocking filter. 
We reduced the data with the SAS v15.0.0 software, selecting single- and double-pixel events ({\sc pattern}$\leq$4) for the pn and single- and multiple-pixel events for the MOS ({\sc pattern}$\leq$12). After removing $\sim1.5$ ks of data affected by high particle-background events in the initial part of the observation, we obtained net exposure times of $\sim50$ ks in the pn  and $\sim55$ ks in the MOS. 

\begin{figure}
\center
		\includegraphics[width=8.0cm]{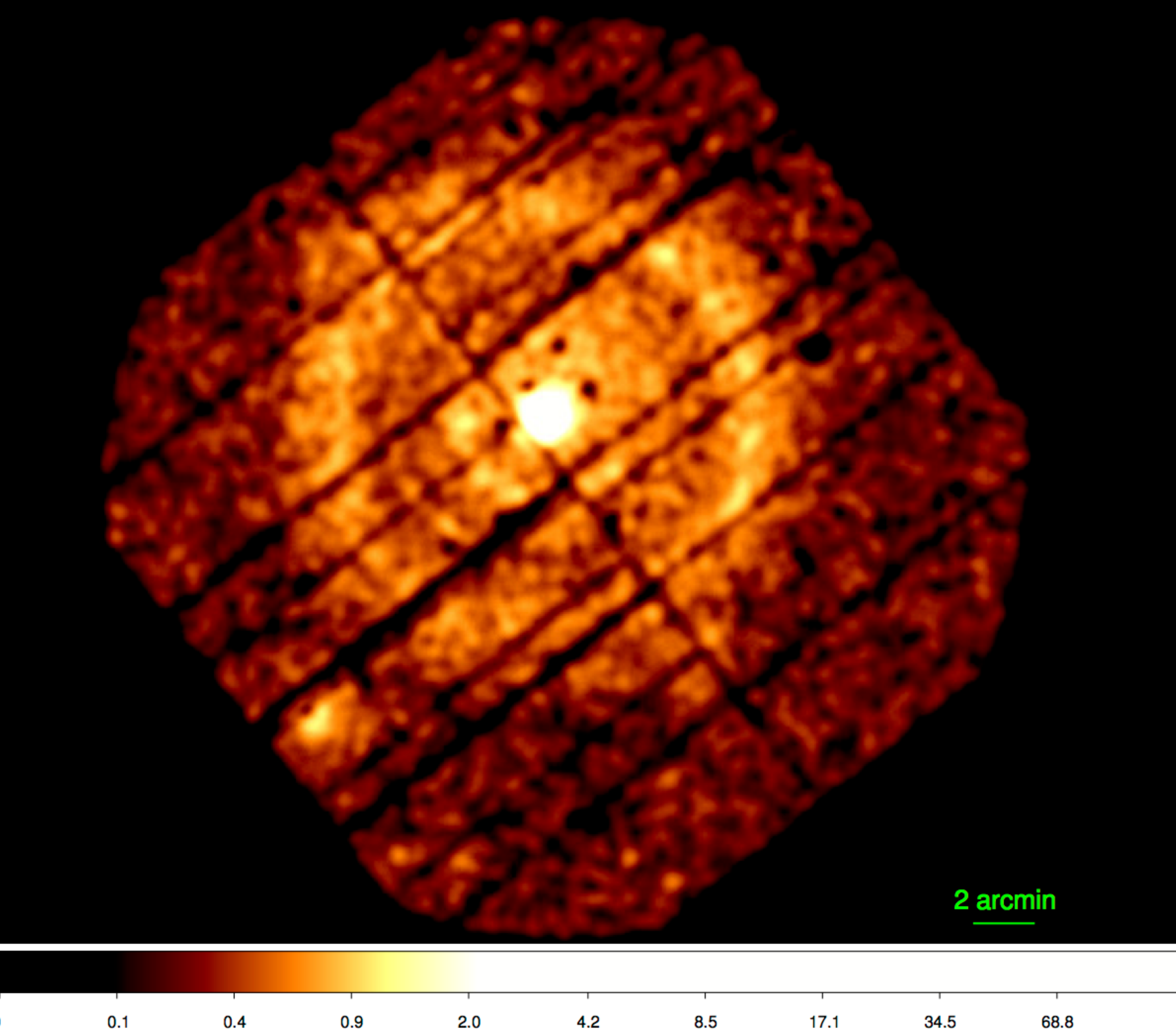}
   \caption{EPIC-pn image (colorbar in counts/pixel) in the 1--2 keV energy band: dust-scattering rings extending up to 9 $'$ are visible around the GRB afterglow. {North is up, East to the left. The galactic center is towards the west side of the image}.}  
        \label{pnimage}
\end{figure}

\begin{figure}
		\includegraphics[width=8.7cm]{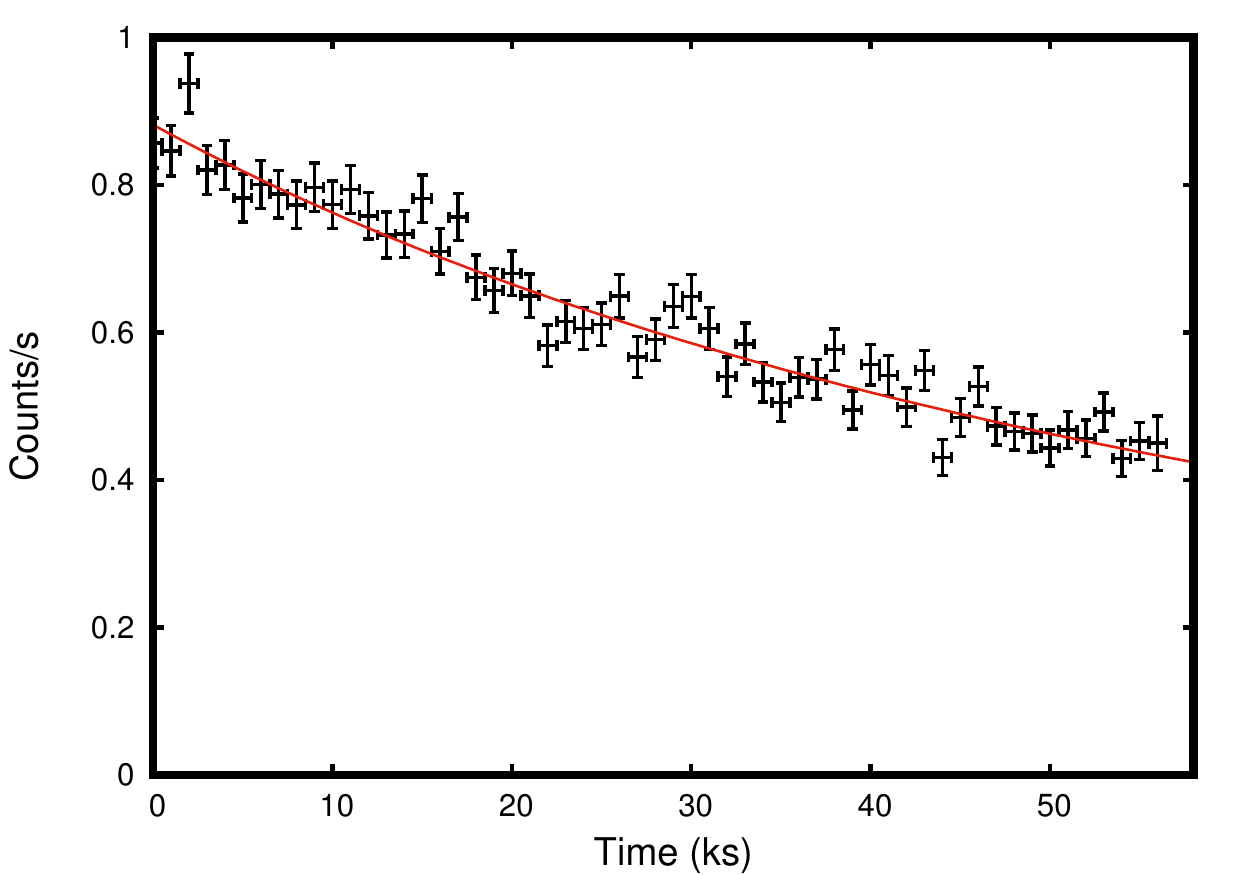}
   \caption{EPIC-pn background-subtracted lightcurve of the GRB afterglow in the energy range 0.3--10 keV sampled with bin size of 1 ks. The decay is modelled by a powerlaw function (red solid line) with index of $-2.11 \pm 0.07$ ($\chi^2/dof=1.2$).}  
        \label{lc-afterglow}
\end{figure}

The image obtained with the pn camera in the 1--2 keV energy range is shown in Fig.~\ref{pnimage}, where a bright afterglow at the GRB position and  the prominent dust scattering rings can be clearly seen.
To derive the properties of the scattering rings, we removed from the event files all the brightest point-like sources in the field-of-view and we removed the out-of-time events\footnote{When \xmm/EPIC is operated in IMAGING mode, out-of-time events are generated when  a photon hits the CCD during the read-out process. This results in the knowledge of the event {\it x} position but not of its {\it y} position because of the readout and shifting of the charges. This effect is much stronger in the pn than in the MOS cameras.} in the EPIC-pn data\footnote{as described in the {\it XMM-Newton} thread https://www.cosmos.esa.int /web/xmm-newton/sas-thread-epic-oot for the case of Full-Frame mode}.

The spectra and lightcurves of the afterglow were extracted from a circular region with radius of 35$''$ centered at the GRB position (R.A. = 315.29859$^{\circ}$, Dec. = 42.221236$^{\circ}$ ) while the background was extracted from a circular region of radius 60$''$ on the same CCD and not containing sources and  scattered emission. 

The spectral fits were carried out with the   {\sc xspec} v12.8.2 software package \citep{arnaud96},  adopting the {\sc tbabs} model for the interstellar absorption with  Solar abundances of \citet{wilms00}. In the following, all the errors in the spectral parameters  are given at the 90\% confidence level for a single interesting parameter.

\section{Data analysis and results}
\label{results}

\subsection{Afterglow}
\label{afterglw}
The light curve of the X-ray afterglow of \src\  is plotted in Fig. \ref{lc-afterglow}, where a smooth count-rate decrease  of about a factor of 2 between the start and the end of the observation can be seen. The decay can be described by a power-law F(t) $\propto$ t$^{\alpha}$ with $\alpha$ = $-2.11\pm0.07$. 

\begin{figure}
		\hspace{-1.7cm}
		\includegraphics[width=6.5cm,angle=270]{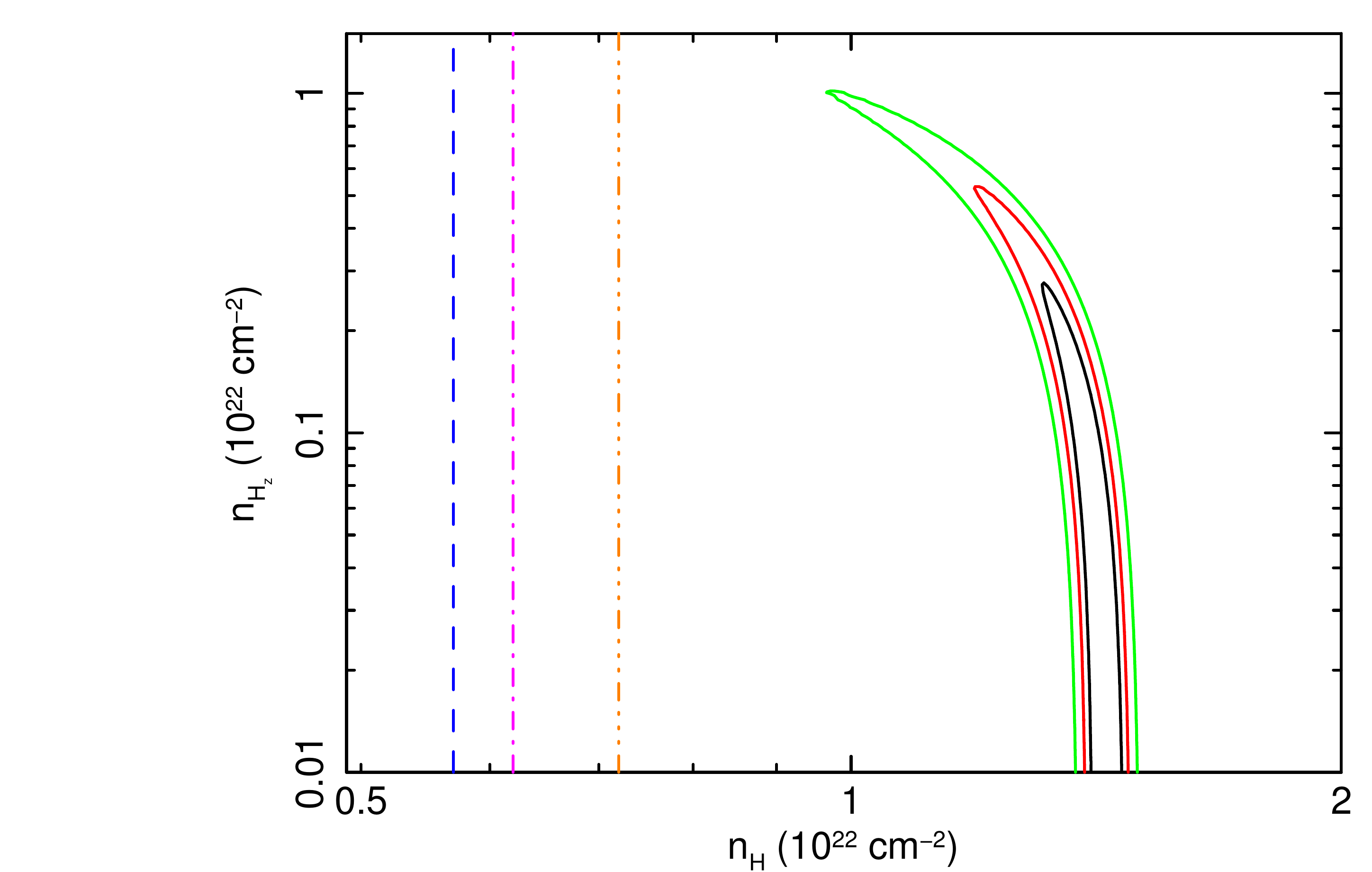}
   \caption{Contour plot, at 1 (black), 2 (red) and 3 (green) $\sigma$ uncertainties, of the Galactic $n$$_\text{H}$ and $n$$_{\text{H}_z}$. The plot shows a significant degeneracy for the two parameters. The blue dashed, the fuchsia dot-dashed and the orange dot-dot-dashed lines represent the Galactic column density expected along the line of sight ($6.2\times10^{21}$ cm$^{-2}$ for \citealt{dickey90}, $5.7\times10^{21}$ cm$^{-2}$ for \citealt{kalberla05} and $7.2\times10^{21}$ cm$^{-2}$ for \citealt{willingale13}).}
        \label{grb_spec_cont}
\end{figure}

The time-averaged spectrum of the afterglow is well described ($\chi^2/dof = 971.05/1036$) by a power-law with photon index  $1.77\pm0.03$, modified  by   a  Galactic plus a host galaxy ($z=0.367$) absorption.
The best fit value of the Galactic absorption  is $n_\text{H} = (1.44_{-0.15}^{+0.04})\times10^{22}$ cm$^{-2}$,  while that of  the host galaxy, $n$$_{\text{H}_z}$, is poorly constrained and with a best fit value significantly lower than $10^{21}$ cm$^{-2}$, as can be seen in Fig.~\ref{grb_spec_cont}.  The absorbed (unabsorbed) 0.3--10 keV flux is $(3.44\pm0.05)\times10^{-12}$ erg cm$^{-2}$ s$^{-1}$ ($(5.9\pm0.1)\times10^{-12}$ erg cm$^{-2}$ s$^{-1}$).
The spectral shape and the column density are consistent with those found with {\it Swift}/XRT  \citep{mingo16}.
We note that the best fit Galactic column density is more than a factor of 2 larger than the total column density expected along this line of sight ($6.2\times10^{21}$ cm$^{-2}$ for \citealt{dickey90}; $5.7\times10^{21}$ cm$^{-2}$ for LAB, \citealt{kalberla05}; $7.2\times10^{21}$ cm$^{-2}$ for \citealt{willingale13}). 

To investigate better this discrepancy, we analyzed an extended source detected by EPIC in the same observation, at coordinates RA $= 21\text{h}01\text{m}51\text{s}$, Dec $= +42^{\circ}03'24''$ , at an angular distance of $\sim12.3$ arcmin from \src . 
The spectral shape and extended nature suggest that this source is a cluster of galaxies.
We fitted its spectrum with an absorbed collisionally-ionized diffuse gas model ({\sc apec} in {\sc xspec}) and found a  temperature of kT$=4.5_{-1.2}^{+2.5}$ keV and a column density of $(7\pm2)\times10^{21}$ cm$^{-2}$, fully consistent with the $n_\text{H}$ expected from \citet{willingale13}.  
With the assumption that the column density in the directions of \src\ and of this source is the same, we fitted the afterglow spectrum with  $n_\text{H}$ fixed to $7.2\times10^{21}$ cm$^{-2}$. A good fit ($\chi^2/dof = 985.33/1037$) was obtained with $n$$_{\text{H}_z} = (1.60\pm0.08)\times10^{22}$ cm$^{-2}$ and a 0.3--10 keV absorbed (unabsorbed) flux of $(2.7\pm0.4)\times10^{-12}$ erg cm$^{-2}$ s$^{-1}$ ($(4.0\pm0.5)\times10^{-12}$ erg cm$^{-2}$ s$^{-1}$).

\begin{figure*}
\center
		\includegraphics[width=17.cm]{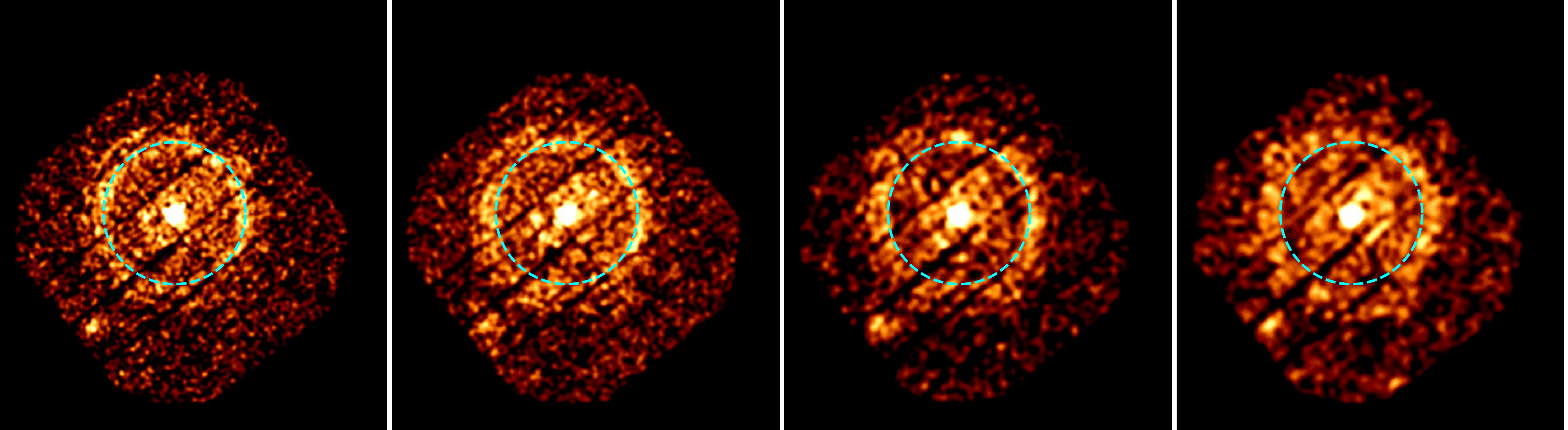}
   \caption{EPIC-pn images in the 1--2 keV energy band, integrated for consecutive intervals of 14 ks starting from the beginning of the  observation. The cyan dashed circle (radius of 6.4 arcmin) is a reference to show better the outer rings expansion. Time increases from left to right.}  
        \label{pnimageint}
\end{figure*}

\begin{figure}
\center
		\includegraphics[width=9.cm]{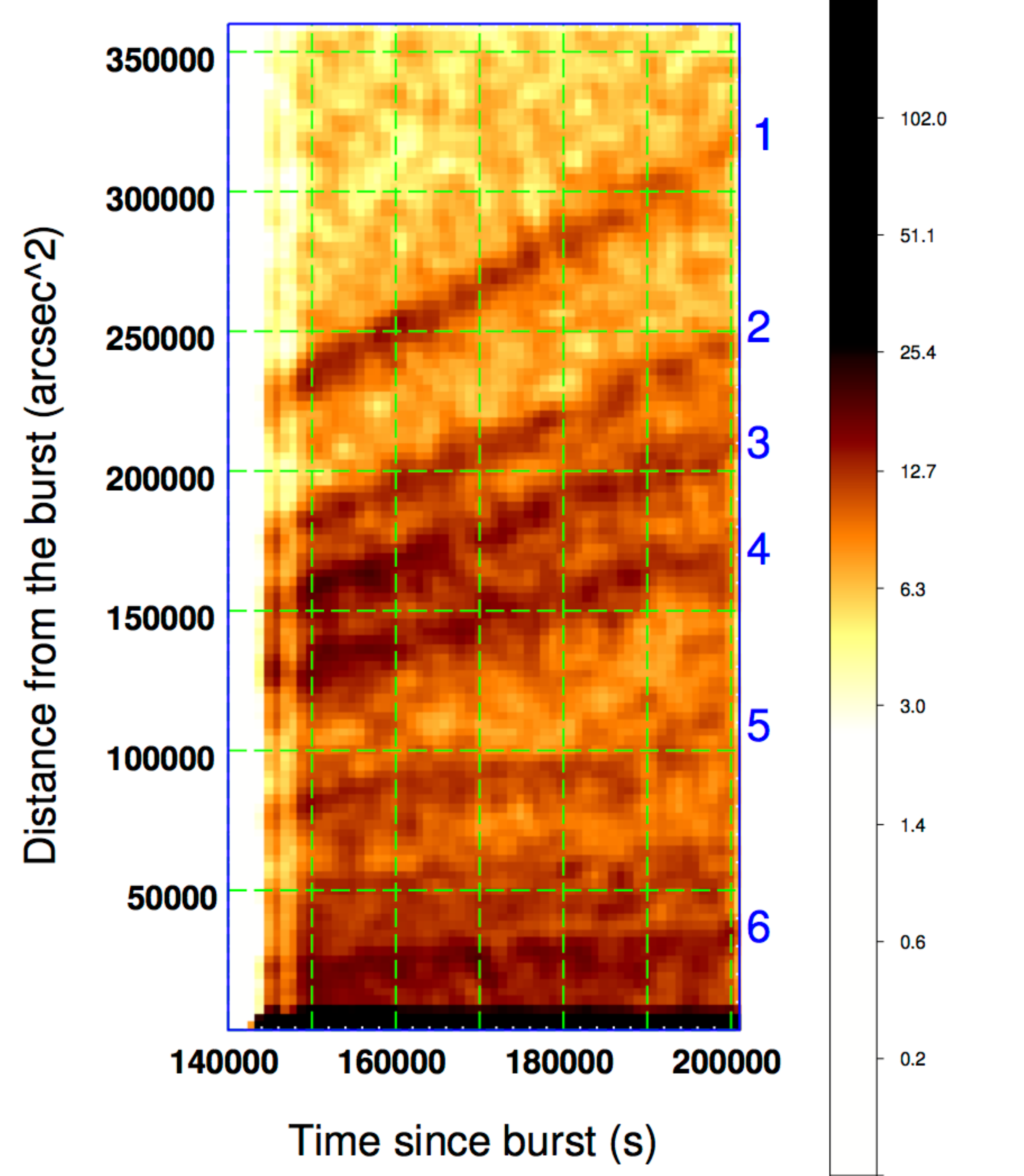}
\caption{Dynamical image (colorbar in counts/pixel) of the EPIC-(pn+MOS) data in the 1--2 keV energy range. The numbers on the right axis label  the different rings.}
        \label{dynamic_img}
\end{figure}

\begin{figure*}
\center
		\includegraphics[width=8.78cm]{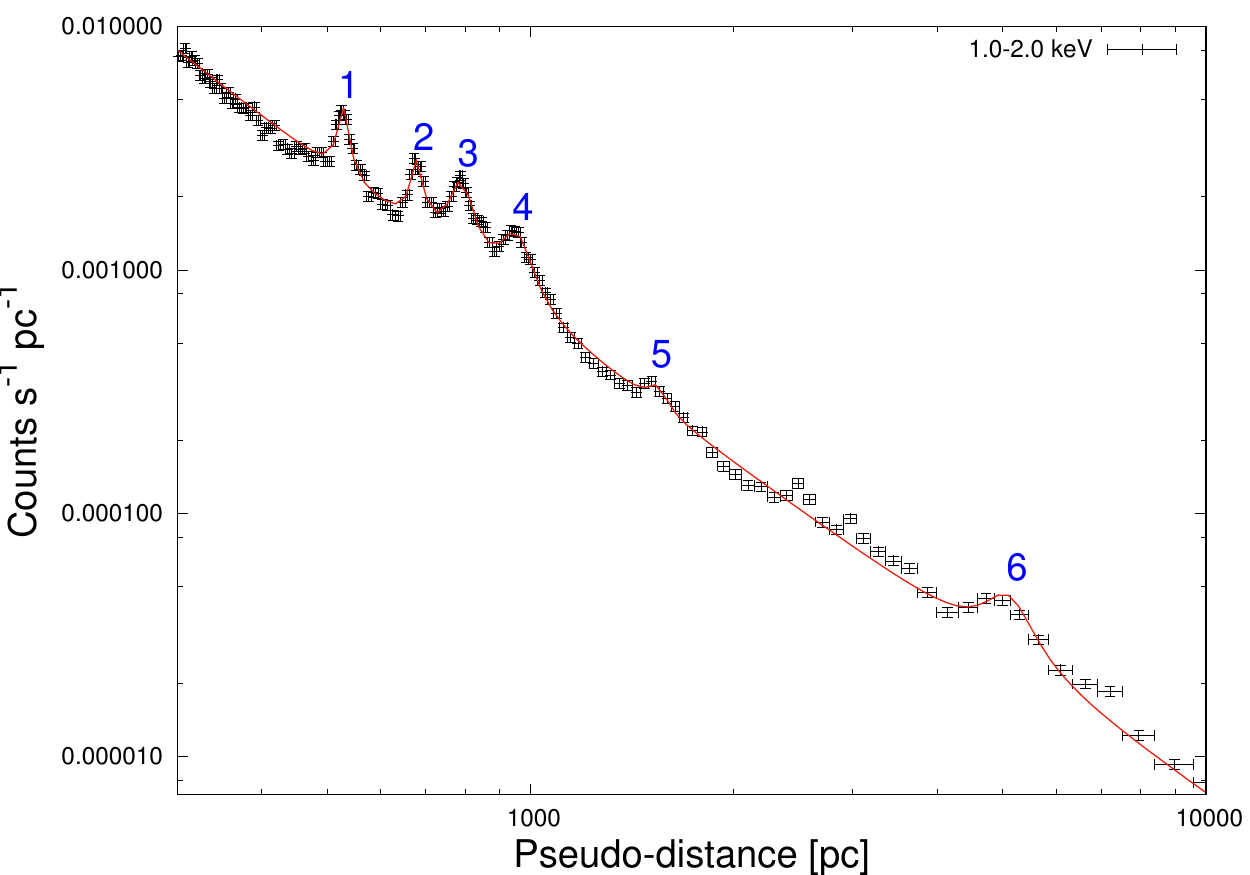}
		\includegraphics[width=8.78cm]{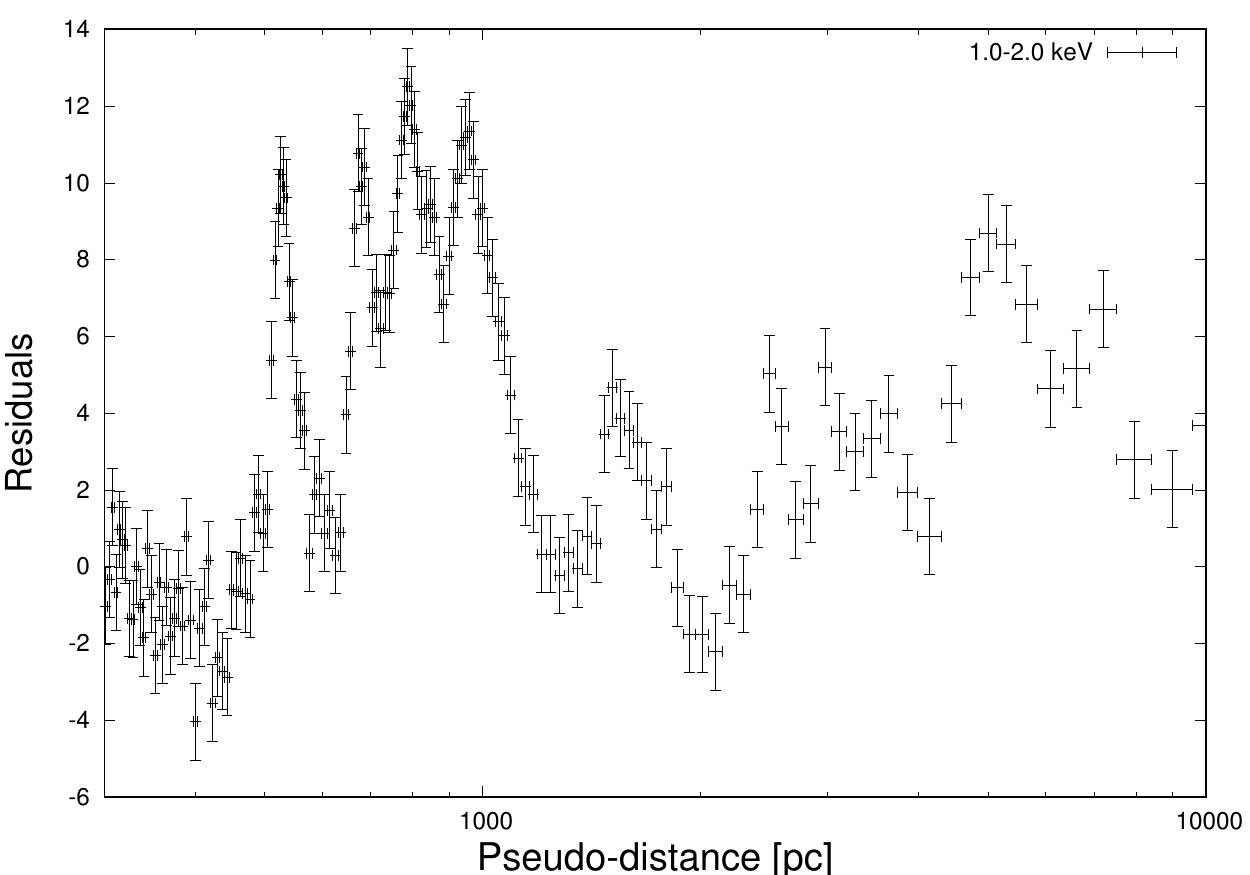}
   \caption{Left-panel: histogram pseudo-distances of the EPIC-pn+MOS instruments (black points) in the 1.0--2.0 keV energy band, fitted with two power-laws plus six lorentzian functions (red solid line). We grouped the data in order to have at least 500 counts in each bin. The numbers, from 1 to 6, represent the label of the ring. Right-panel: residuals (in units of $\sigma$) of the best fit with two powerlaws with index $-1.91\pm0.08$ and $-2.87\pm0.20$. \newline
In both plots, six strong features are clearly visible at distances of $\sim$530 pc, 680 pc, 790 pc, 950 pc, 1540 pc and 5080 pc.}  
        \label{dynamic_img_pseudo}
\end{figure*}

\subsection{Dust-rings}

To illustrate the angular expansion of the rings, we created four EPIC-pn images in the 1--2 keV energy band, corresponding to consecutive  time intervals of $\sim$14 ks each (Figure~\ref{pnimageint}). They clearly show the presence of at least three expanding rings, with average radii of $\sim3.2$ arcmin, $\sim$5.2 arcmin and  $\sim$7.3 arcmin. The outer ring is broader than the others and shows the most prominent expansion. Describing the angular  expansion as $\theta=K (t-\text{T}_0)^{1/2}$,  with $\theta$   in arcmin and ($t-\text{T}_0$) in days ($\text{T}_0$ is the GRB time and $t$ is the mean time of the \xmm\ observation), we estimated $K\sim2.3$, 3.7, and 5.2 arcmin/day$^{0.5}$ for the three rings. Using equation 2, these expansion coefficients correspond to dust distances of $\sim4$ kpc, $\sim1.5$ kpc and $\sim0.7$ kpc, respectively. Therefore, the outermost ring is compatible with being the one detected by {\it Swift}/XRT at earlier times and for which  \citet{tiengo16} estimated  d$\sim$0.8 kpc. 

To increase the sensitivity for the detection of faint expanding rings not easily seen in the time integrated images of Fig.~\ref{pnimageint}, we used the method based on so called  {\it dynamical images} first introduced by \citet{tiengo06}. A dynamical image is created by converting the detector position ($x_i$ and $y_i$) and time of arrival ($T_i$) of each event into a new set of coordinates given by:
\begin{gather}
\begin{flalign*}
\nonumber & t_i = T_i - \text{T}_0  &\\ 
\nonumber & \theta_i^2 = (x_i - X_\mathrm{GRB})^2 + (y_i - Y_\mathrm{GRB})^2 &
\end{flalign*} 
\end{gather}
where $\text{T}_0$, $X_\mathrm{GRB}$ and $Y_\mathrm{GRB}$ are the GRB start time and  spatial coordinates  (the latter are derived from the afterglow position in this observation). In an image based on these coordinates, a (non-variable) source  appears as a horizontal line at constant $\theta^2$,   while an expanding ring centered at $X_{grb}$, $Y_{grb}$ appears as an inclined line with angular coefficient proportional to the  distance of the dust-scattering layer.

We first performed this analysis separately on the data of the three EPIC cameras and did not find any significant difference. We then stacked the  pn and  MOS events to increase the signal-to-noise ratio and hereafter, unless explicitly stated, we shall always refer to the combination of the EPIC data.
In Fig.~\ref{dynamic_img}, we show the {\it dynamical image} in the 1--2 keV energy range, where at least six inclined lines associated to the dust-scattering rings are visible. We note that with this method we can identify more rings than in a simple study of the radial profile of the EPIC sky image.

Following \citet{tiengo06} and, for each event, we can define a {\it pseudo-distance} $D_i$ as:
\begin{gather}
\begin{flalign*}
& D_i = 2ct_i / \theta_i^2 = 827 t_i [\text{s}] / \theta_i^2 [\text{arcsec}] \text{  }pc.  &
\end{flalign*} 
\end{gather}
We called it {\it pseudo-distance} as it assigns a distance to both background and source/expanding ring events. Clearly, for background events, this value is not a real distance. On the other hand, events coming from an expanding ring cluster around a specific value of  $D$ which corresponds to the ``true'' dust-layer distance.  
Hence, to derive quantitatively the properties of the rings, we created a histogram of the pseudo-distances (Fig.~\ref{dynamic_img_pseudo}-left), corrected for the exposure time. 
In this histogram, the background counts form a smoothly decaying continuum while the rings appear as individual peaks.
A uniform background should be described by a power-law with index --2. On the other hand, the background component due to the diffuse X-ray emission is subject to telescope vignetting effects, which produce a significant deviation from the expected power-law.
Indeed, we found that the continuum shape can be described with the sum of two power-laws with indexes of $-1.91\pm0.08$ and $-2.87\pm0.20$. The peaks due to the rings are instead well described by Lorentzian functions. { By fitting the pseudo-distance histograms with these models, we can estimate the statistical significance of the rings seen in the dynamical images: we found that the addition of {\it each} of the six Lorentzian models significantly improves the reduced $\chi^2$, with an F-test probability that the improvement is not obtained by chance $\gg$$5\sigma$ (for three additional degree of freedom).} The histogram in the 1--2 keV energy range and the best fit with two power-laws and six Lorentzians are shown in Figure~\ref{dynamic_img_pseudo}-left, while in Figure~\ref{dynamic_img_pseudo}-right we show the residuals (in units of $\sigma$) of the fit with only  the continuum. The best fit centroids of the Lorentzians give accurate measurements of the distances of the  dust-layers, which range from $528.1\pm1.2$ pc to  $5079\pm64$ pc (see Table~\ref{table_spectra} for all the values).  

We estimated, through simulations using the EPIC Point Spread Function\footnote{We used the analytical model of  Ghizzardi (2002, XMM-SOC-CAL-TN-0029), which accurately describes the spatial and energy dependence of the EPIC-PSF, but not its azimuthal structure (see \citealt{read11}).} (PSF), {that the best fit FWHMs of the two innermost rings are slightly larger than the expected instrumental broadening (although consistent within  2$\sigma$ uncertainty) while those of  the other rings are instead significantly larger.} The corresponding  thicknesses of the  dust-layers  are given in Table~\ref{table_spectra}.

Finally, we also note in Fig.~\ref{dynamic_img_pseudo} the presence of two other, weaker peaks at $\sim2.5$ kpc and $\sim3$ kpc. We excluded them from the analysis because their properties could be only poorly constrained.

\subsection{Spectral analysis of the rings}
\label{tail_spec}

A standard   extraction of the spectra of the rings  (i.e.  selecting the counts from annular regions  in the EPIC images)  presents some critical issues. In fact, due to  their expansion during the observation, the rings spatially overlap.  The selection of fixed annular regions would yield spectra with mutual contamination of the adjacent rings. In addition, it would be difficult to select background regions because most of the field of view is covered by the rings. 
The best approach to overcome  these problems  is to extract directly the background-subtracted spectra  by integrating the Lorentzian functions fitted to the histograms of the pseudo-distances  for  different energy bins.

\begin{figure}
\center
		\includegraphics[width=5.8cm,angle=270]{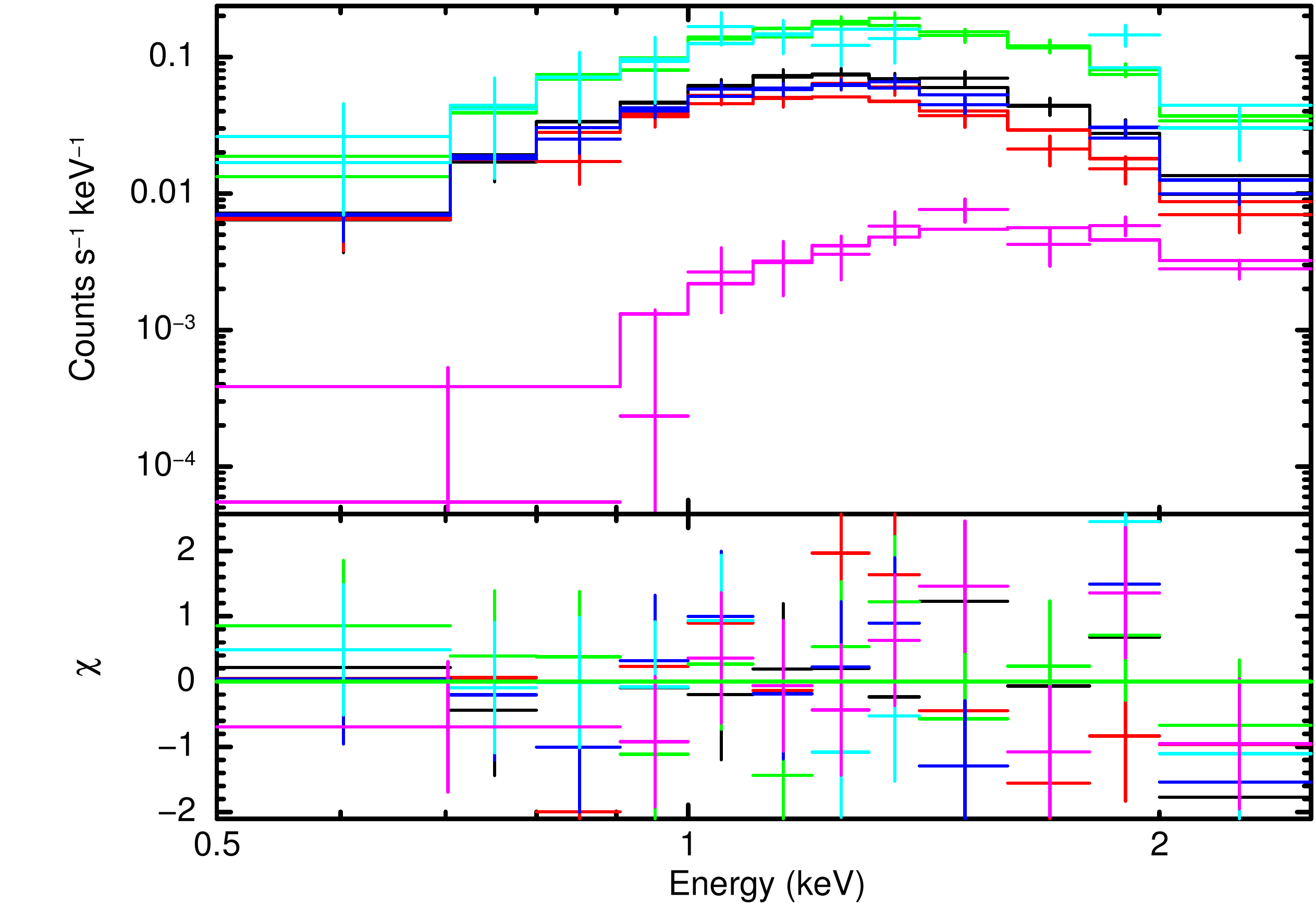}
   \caption{Spectra of the dust-scattering rings (top-panel) fitted with a {\sc tbabs*constant*dust-model}   model and the corresponding residuals (bottom panel). The black, red, green, blue, cyan and purple points represent data from ring 1, ring 2, ring 3, ring 4, ring 5 and ring 6, respectively. Because of a strong background contamination, some spectral points were removed around the energy of the instrumental Al feature ($\sim$1.5 keV) for rings 4 and 5.}  
        \label{ring_spectra}
\end{figure}

The rings are spectrally soft and  detected with the highest significance in the  0.5--2.5 keV range. We divided this range in different bins and for each one  we created  the histogram of the pseudo-distances. We fixed the slopes of the continuum power-laws, as well as the centroids and   widths of the Lorentzians,  to their values obtained in  the total 0.5--2.5 keV histogram; we then fitted only the  normalizations of the power-laws and of the Lorentzians in each energy bin.
By integrating the best fit Lorentzian functions, we obtained the background-subtracted ring spectra shown in Figure~\ref{ring_spectra}. They clearly show that rings from 1 to 5 have  similar fluxes (rings 3 and 5 are the brightest), while ring 6 is significantly fainter but its spectrum is the hardest one. 

The intensity of each ring as a function of time and energy depends on the spectrum and fluence of the burst  and on the optical depth of the corresponding dust layer.
To derive the latter quantity from a fit of the ring spectra, we followed the procedure outlined in \citet{shao07}.
We adopted the  grain composition and size distribution of the \textit{BARE-GR-B} model\footnote{This model includes polycyclic aromatic hydrocarbons (PAH), silicate grains and  bare graphite grains with the abundances of B-type stars. See Fig.14 of \citet{zubko04} for a comparison with the dust-distribution from \citet{mathis77}} presented in \citet{zubko04}, which was found to describe quite well the properties of dust-layers in the Galaxy (see e.g. \citealt{tiengo10,pintore17,Jin17}). To compute the expected spectrum of each ring at the time of the \xmm\ observation, we used the Rayleigh-Gans approximation for the scattering cross section, the dust distances derived above, and the burst fluence obtained by extrapolating the Konus-Wind spectrum ($1.2\times10^{-5}$ erg cm$^{-2}$ in the 0.3--10 keV range).
The resulting model, which we  implemented as a  {\sc table model}  in XSPEC,  has as free parameter only the normalization, which gives the dust column density $n_{\text{H}_{dust}}$ of the scattering layer. For each spectrum, we created response matrices and ancillary files (for extended sources) for an annular region with inner and outer radii defined by the region swept by the ring centroids from the start to the stop time of the observation. {The arf files were created by taking into account also the exposure map in the selected annular regions.}
In the range 1.3--1.6 keV, the background continuum was significantly affected by the instrumental Al feature (at 1.49 keV) and such a contamination was particularly prominent for rings 4 and 5. Due to the resulting  poor characterization of the continuum, we excluded this energy bin from the spectra of rings 4 and 5. 

\begin{table}
\center
          \caption{Column density of the interstellar medium and of the dust layers, and their corresponding unabsorbed fluxes in the 0.3--10 keV energy range, as estimated by the fits of the ring spectra. Errors at $90\%$  confidence level.} 
      \label{table_spectra}
\scalebox{0.95}{\begin{minipage}{24cm}
\begin{tabular}{lcccc}
\hline
&  $n_{\text{H}_{dust}}$ & Flux & Distance & Width \\
&  ($10^{21}$ cm$^{-2}$) & erg cm$^{-2}$ s$^{-1}$ & (pc) & (pc)\\
\hline
Ring 1 &  $6.9^{+0.7}_{-0.6}$ & 1.1$\times10^{-12}$ & $528.1 \pm 1.2$& $23.4 \pm 3.3$ \\
Ring 2 &  $4.5^{+0.5}_{-0.4}$ & 0.66$\times10^{-12}$ & $679.2 \pm 1.9$ & $32.3 \pm 5.7$\\
Ring 3 &  $14.6^{+1.2}_{-1.1}$ & 2.1$\times10^{-12}$ & $789.0 \pm 2.8$ & $75 \pm 10$\\
Ring 4 &  $5.0^{+0.4}_{-0.4}$ & 0.66$\times10^{-12}$ & $952 \pm 5$ & $116 \pm 15$\\
Ring 5 &  $11.4^{+2.1}_{-2.1}$ & 1.2$\times10^{-12}$ & $1539 \pm 20$& $106 \pm 60$ \\
Ring 6 &  $0.7^{+0.1}_{-0.1}$ & 0.045$\times10^{-12}$ & $5079 \pm 64$ & $1000 \pm 400$\\
\hline

\end{tabular}
\end{minipage}}
\end{table}

We fitted simultaneously the ring spectra including Galactic absorption with   n$_{\text{H}}$ 
 linked to a common value ({\sc tbabs$\times${dust}} in {\sc xspec}). 
We found a good fit ($\chi^2/dof = 67.46/54$) with n$_{\text{H}}=(1.04\pm0.06)\times10^{22}$ cm$^{-2}$ and the dust-column density values for the dust-layers shown in Table~\ref{table_spectra}.
They are in the range $\sim0.07-1.5\times10^{22}$ cm$^{-2}$ and give  a total dust column density of $\sim4.3\times10^{22}$ cm$^{-2}$. This is larger than the value of the Galactic column density expected along the line of sight ($7.2\times10^{21}$ cm$^{-2}$, \citealt{willingale13}) and the value ($1.44\times10^{22}$ cm$^{-2}$) derived from our fit to the afterglow spectrum with Galactic and host absorption as free parameters. 
Such a discrepancy may be explained considering that values of the dust grain-to-hydrogen atom ratio larger than 1 have been  observed also in other Galactic clouds \citep[see e.g.][]{vuong03,hasenberger16,pintore17}. The difference is likely due to the fact that the gas-to-dust ratio is not homogenous in the Galaxy, while the normalization of the dust model we adopted here is based on an average value of this ratio. Alternatively, one is forced to conclude that the soft X-ray burst fluence we derived by extrapolating the Konus-WIND spectrum is actually underestimated {by at least of factor of $\sim6 (3)$ to be in accordance with the column density of $7.2\times10^{21}$ cm$^{-2}$ ($1.44\times10^{22}$ cm$^{-2}$)}.

{Finally we note that the adopted Raylegh-Gans approximation, used to calculate the scattering cross-section, assumes that photon wavelengths is much larger than the grain size, the shape of the grain is spherical, the photon absorption is negligible and the scattering angle is small \citep[e.g.][]{smith98}. For the expected sizes of the interstellar dust grains, such assumptions are generally valid for X-ray photons $\geq 1$ keV. For this reason, we also checked the goodness of such approximation in our analysis and we fitted the spectra for energies $>1$ keV, finding that the dust-column densities are fully consistent with those inferred in the broader energy range. This implies that, although the Rayleigh-Gans approximation may not be fully satisfied, it does not significantly affect our results. }

\section{Discussion}
\label{discussion}

The good imaging capabilities and high sensitivity  of the \xmm/EPIC detectors, and the use of an ad-hoc analysis technique based on ``dynamical-images'', allowed us to find evidence of at least six expanding rings around the position of \src .

\begin{figure}
\center
		\includegraphics[width=8.4cm]{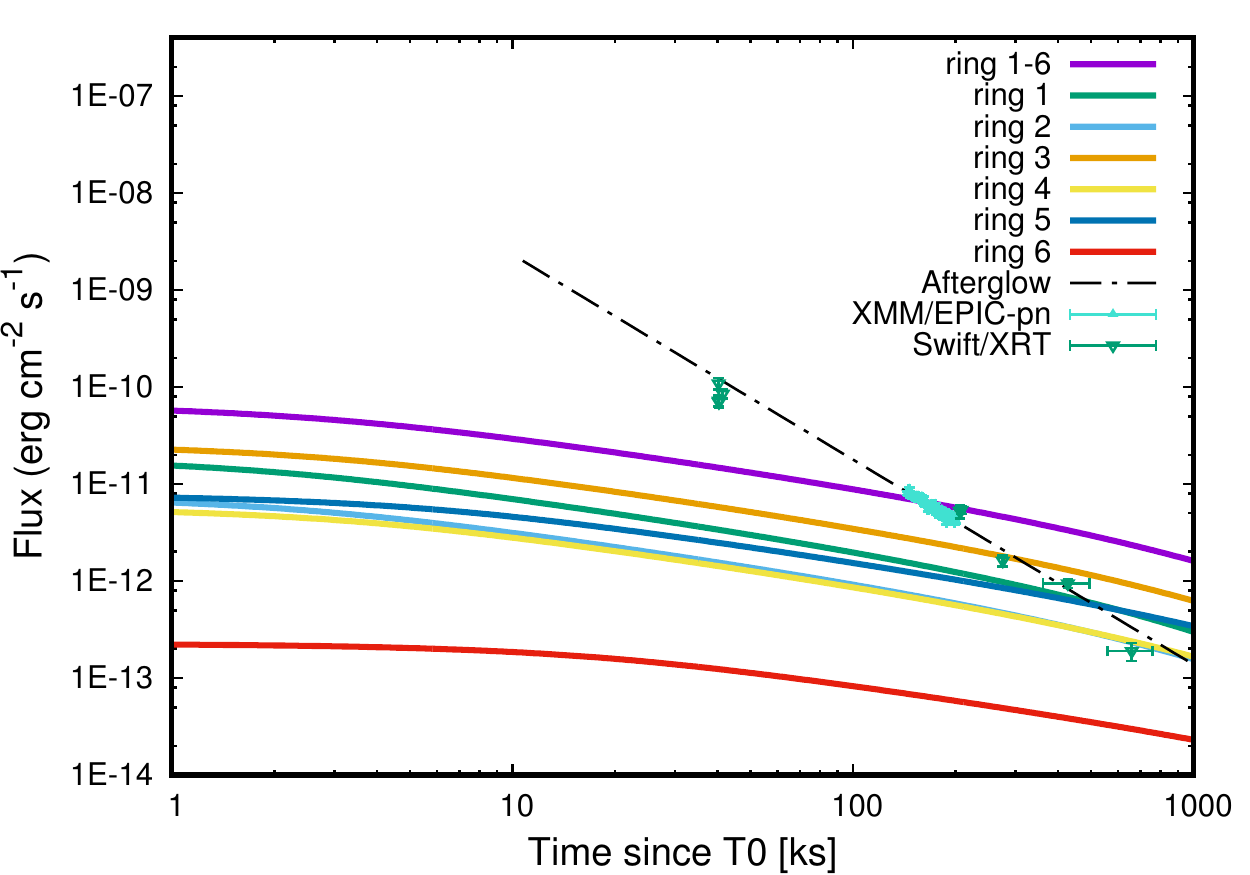}\vspace{-0.5cm}
\caption{Unabsorbed 0.3--10 keV flux of the \xmm/EPIC-pn (cyan filled triangles) and {\it Swift/XRT} observations (green filled reversed triangles) of the \src\ afterglow. The black dashed line shows the flux extrapolated from the best fit powerlaw of the \xmm\ EPIC-pn data.
The solid lines represent instead the dust-model lightcurves. The total contribution of the dust-scattering flux decays less quickly than the afterglow and dominates the emission starting $\sim100$ ks after the GRB.}
\label{dust_lc}
\end{figure}

\begin{figure}
\center
		\includegraphics[width=8.8cm]{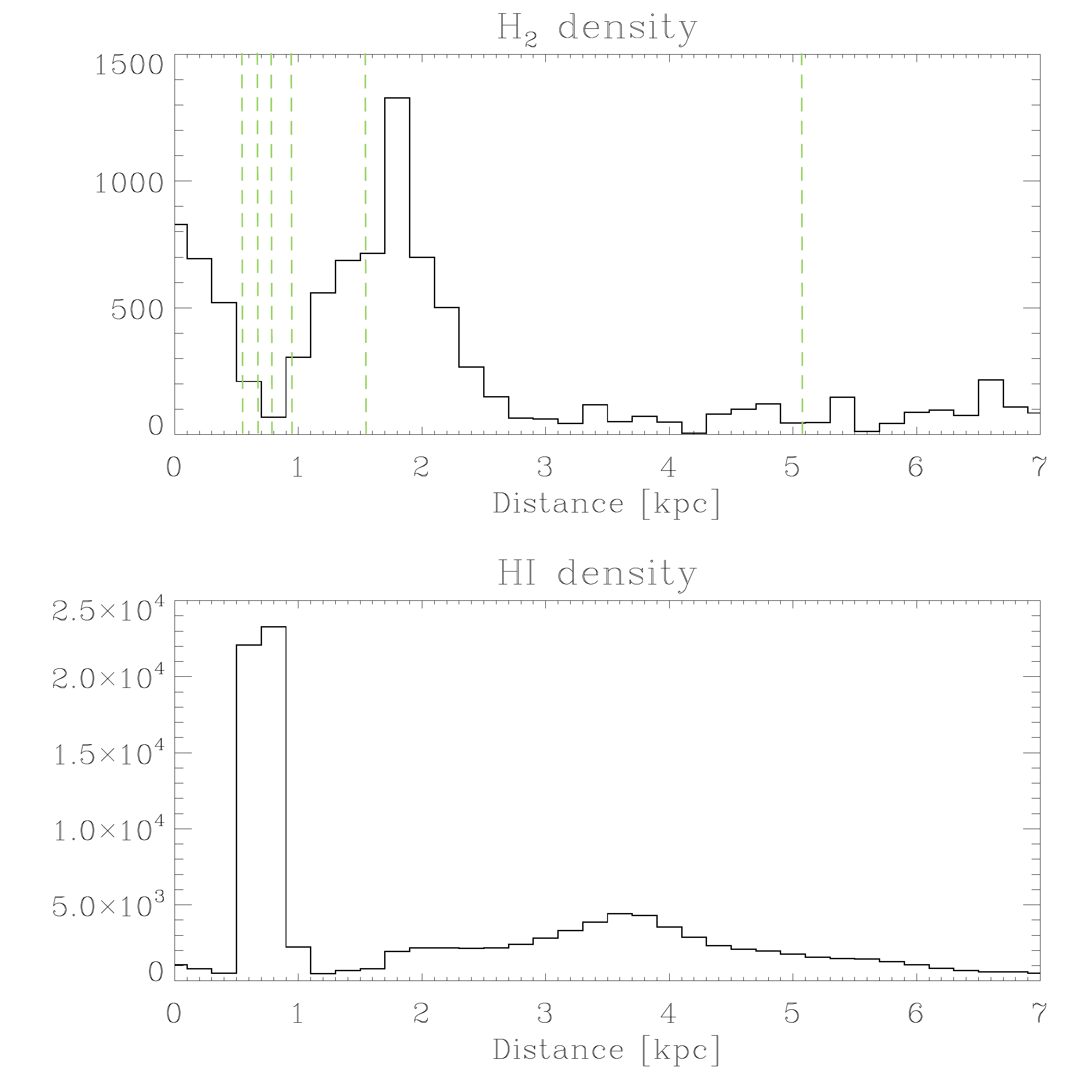}\vspace{-0.5cm}
\caption{Galactic density profiles in unity of $10^{4}$ atoms cm$^{-3}$ of the H I and H$_2$ clouds towards the direction of the GRB. In each step of the histogram, the uncertainties on the densities are 20$\%$ and 40$\%$ for H$_2$ and H I. The dashed lines indicate the position of the dust-layers found in this work.}
        \label{himap}
\end{figure}

\begin{figure*}
\center
		\includegraphics[width=5.85cm]{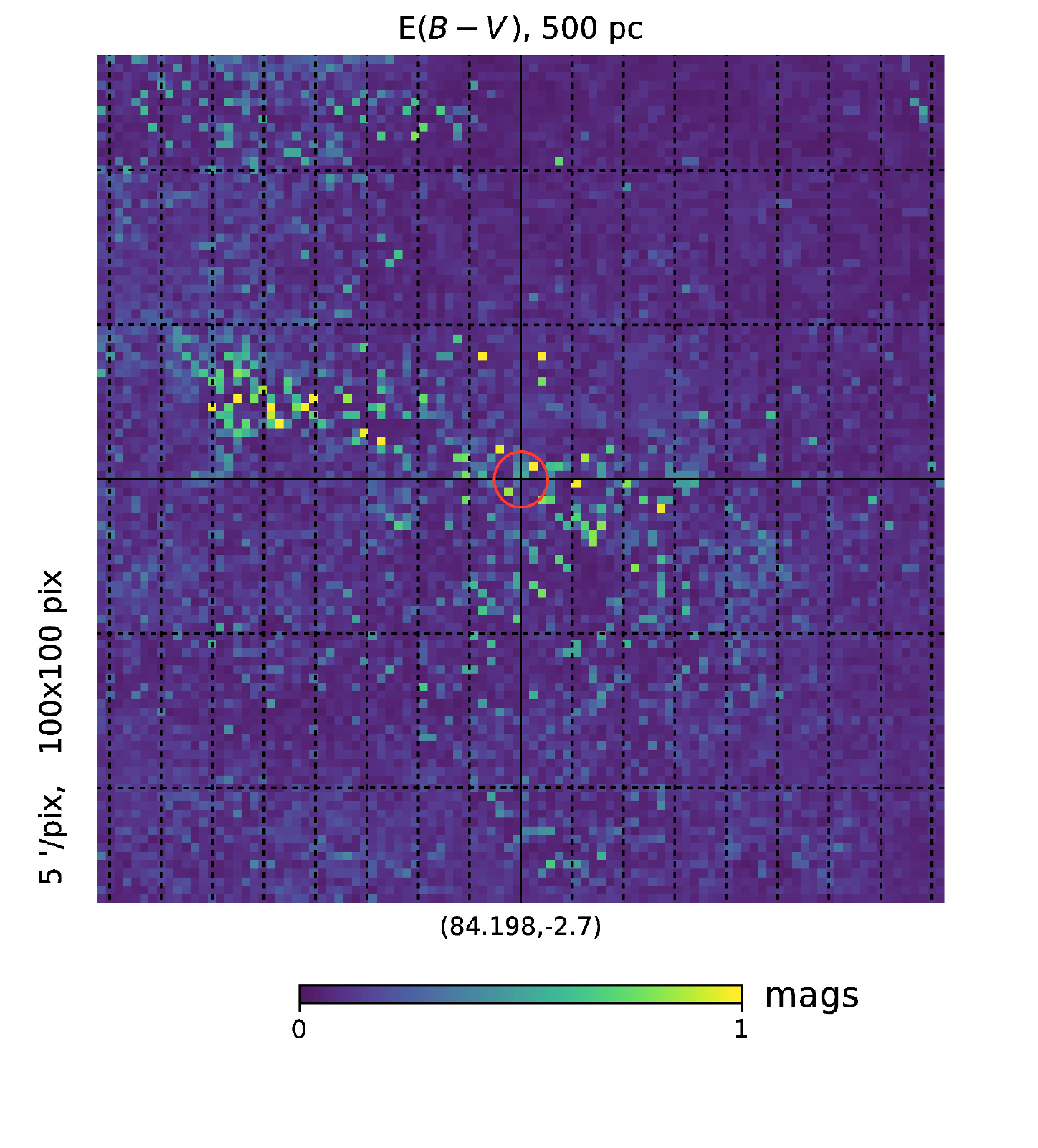}
		\includegraphics[width=5.85cm]{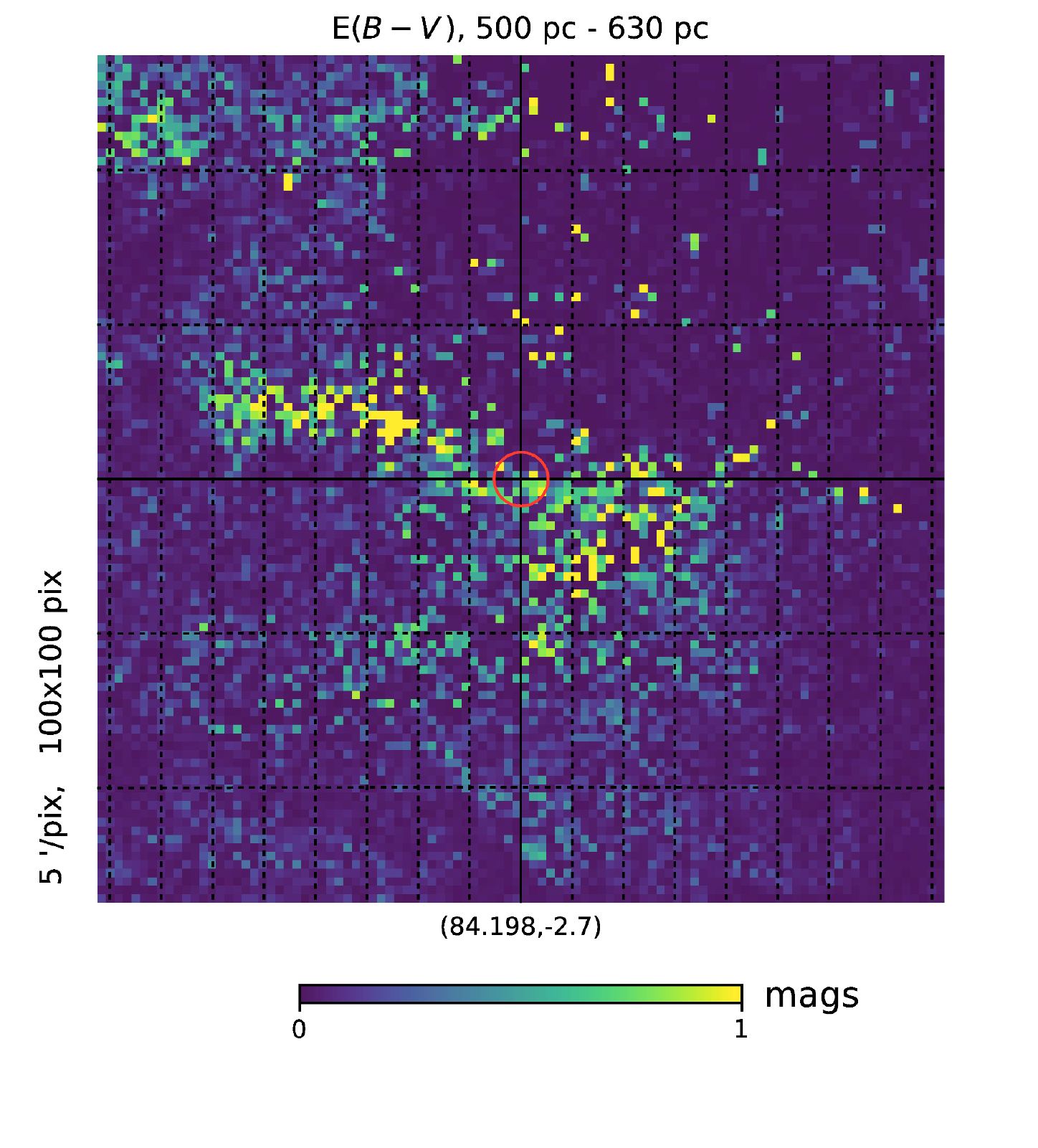}
		\includegraphics[width=5.85cm]{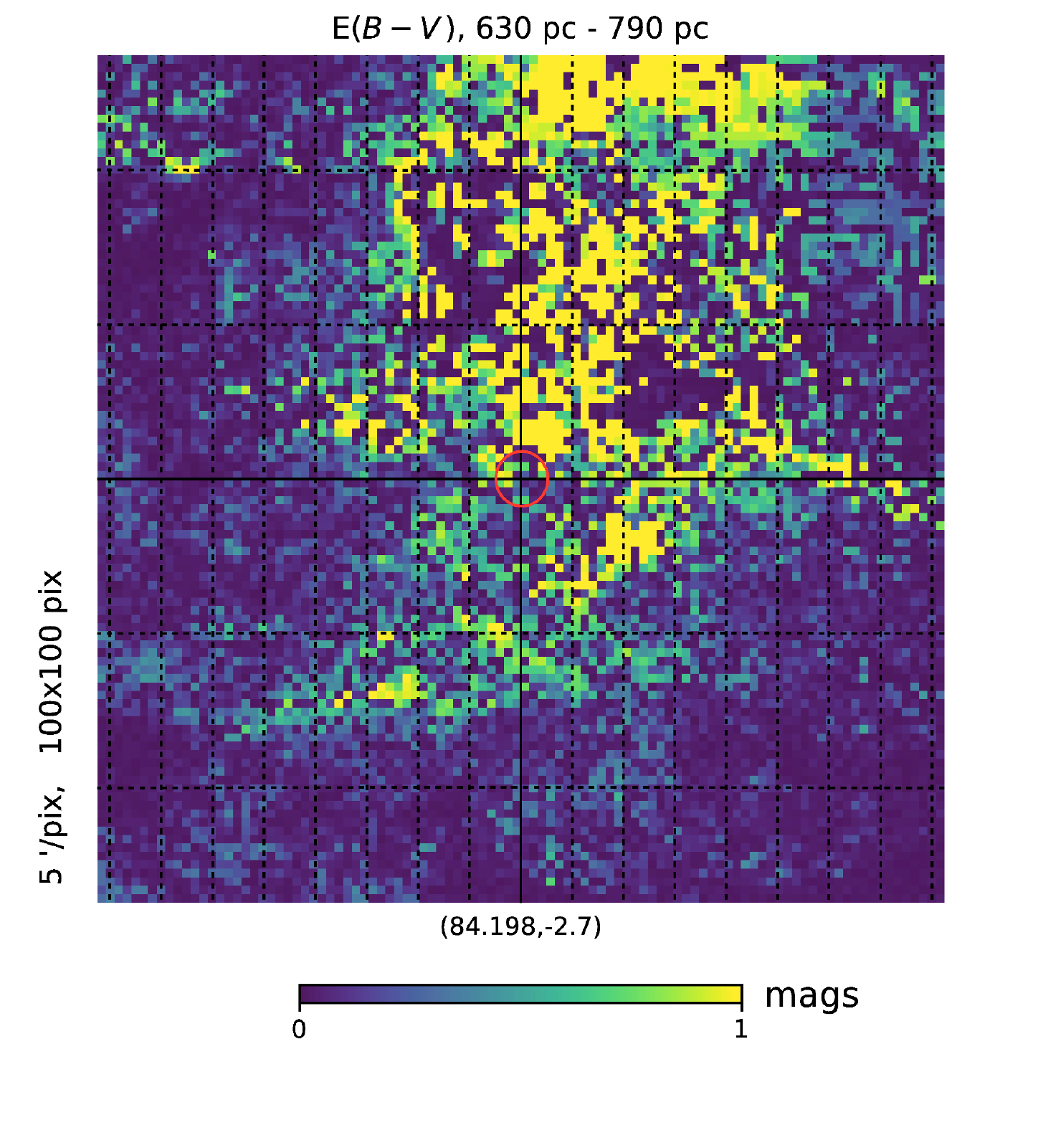}
		\includegraphics[width=5.85cm]{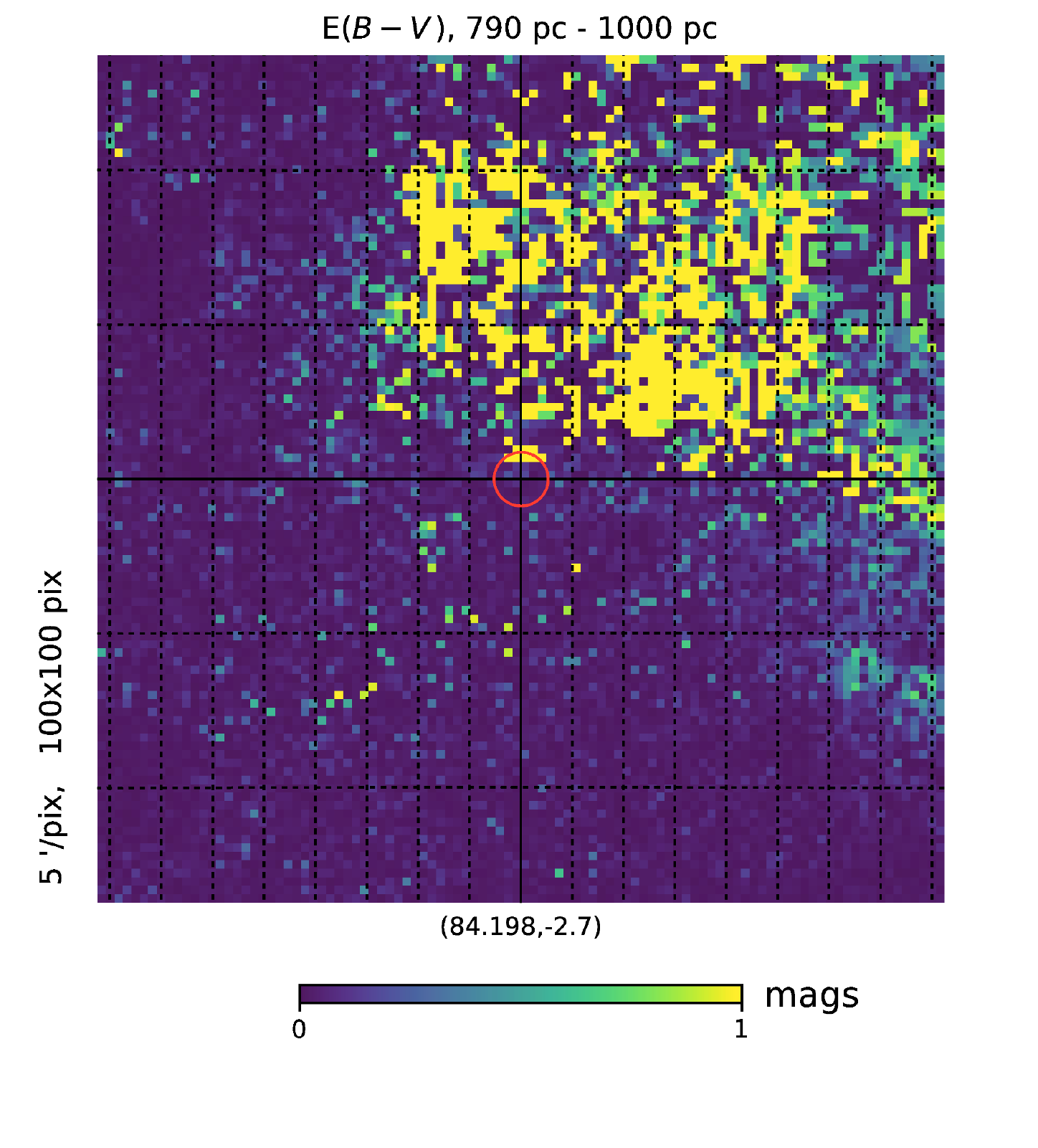}
		\includegraphics[width=5.85cm]{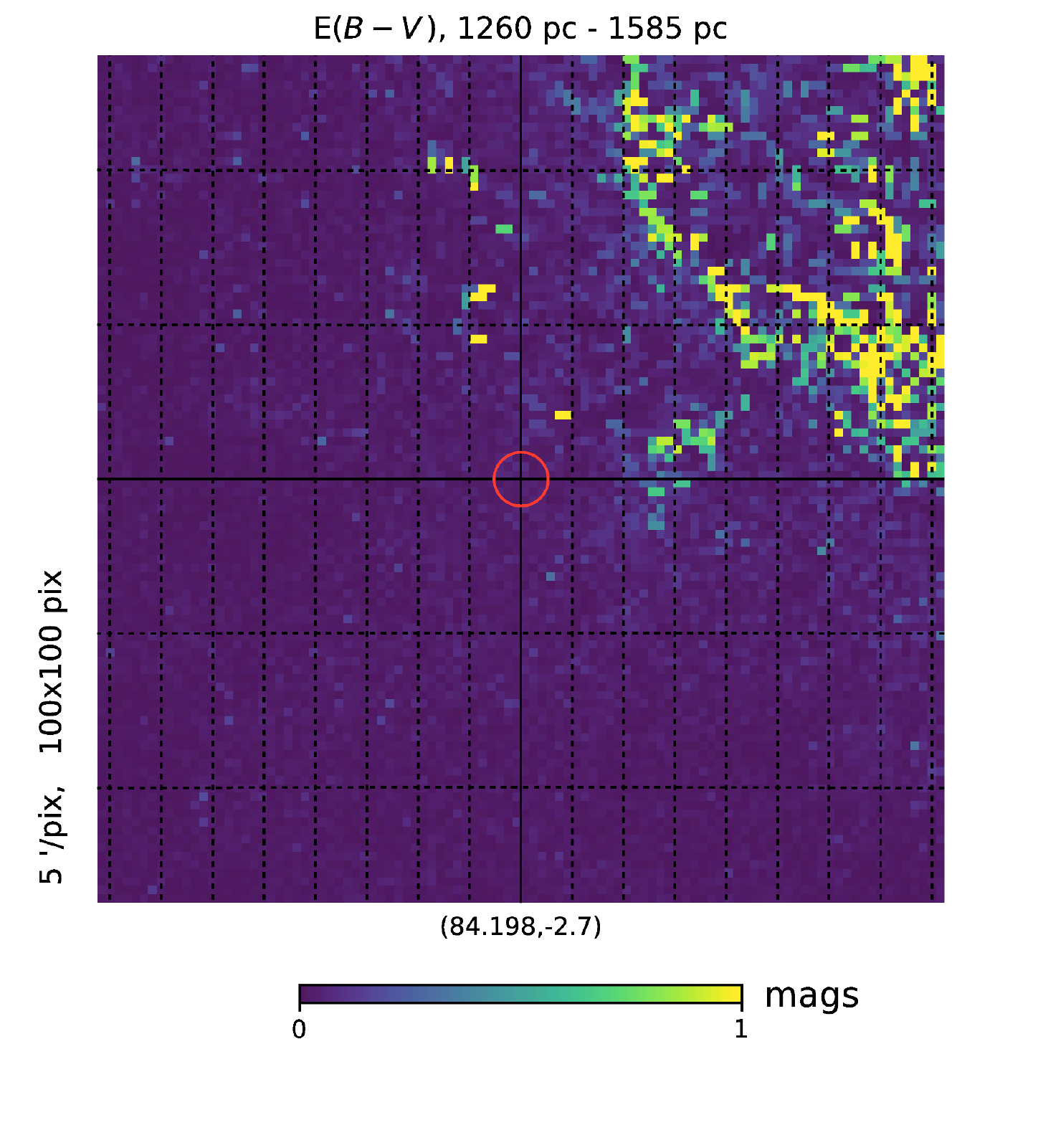}
		\includegraphics[width=5.85cm]{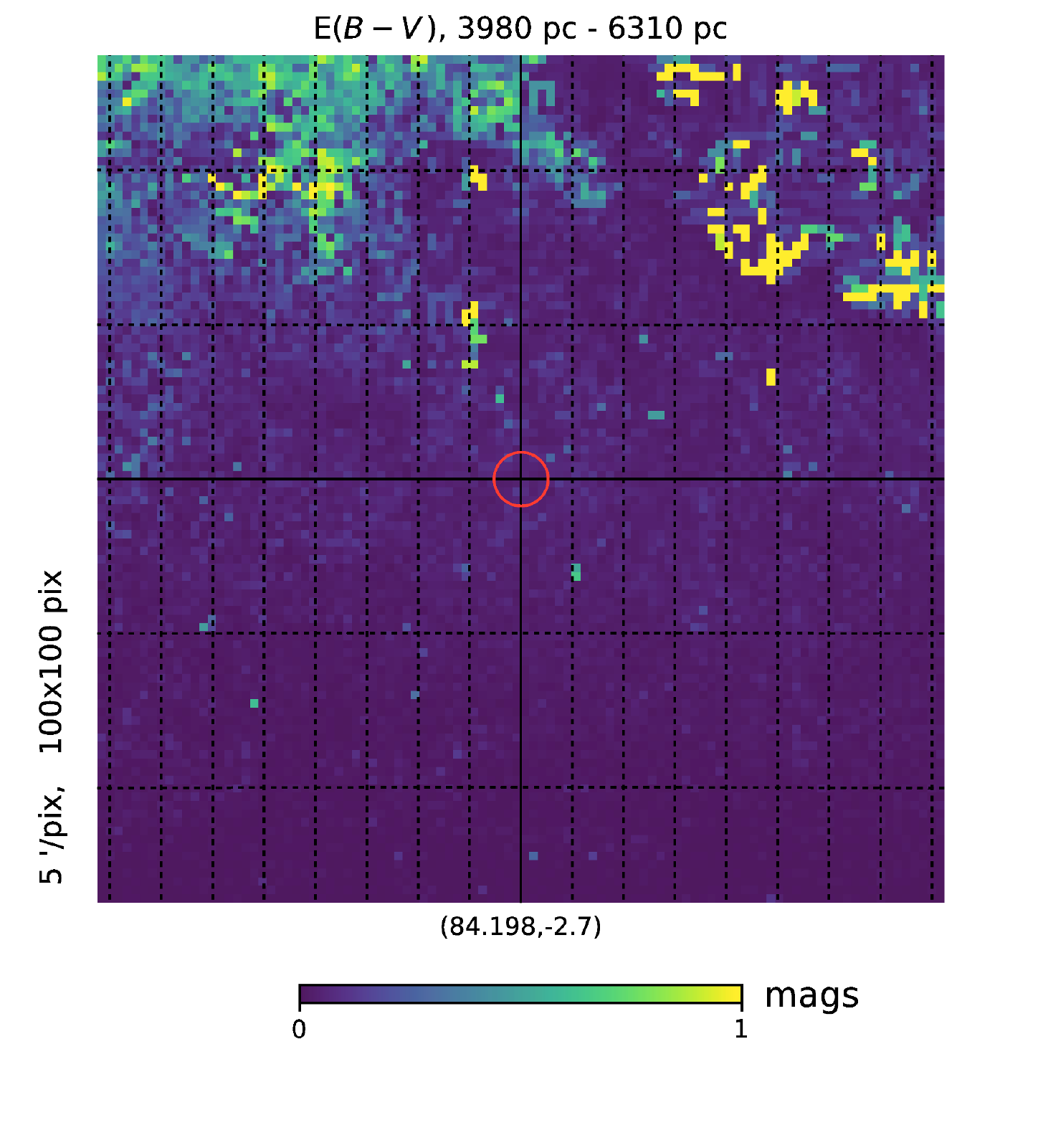}
\caption{Pan-STARSS 1 extinction maps along the line of sight of \src (Galactic coordinates l=84.2°, b=-2,7°). Each pixel in the map is 5 arcmin, for a total of 100$\times100$ pixels, and the red circle represents the \xmm\ field of view. The maps are evaluated at the distances found for the rings: the largest extinction is placed in the range 630--790 pc, i.e. at the position of ring 3, which is also the one with the highest column density.}
        \label{pannstar}
\end{figure*}

The expansion rate of the rings allowed us to derive a dynamical measure of the dust-layers distances (see Table~\ref{table_spectra}).
Remarkably, the distances of the dust-layers are constrained with a precision of less than 1$\%$ just basing the analysis on simple geometrical considerations, which makes this the best characterization by far of the Galactic clouds along this line of sight. 
The {\it Swift}/XRT data, taken 40 ks after the GRB revealed a halo produced by dust at a distance of $\sim$800 pc \citep{tiengo16}, that we can associate with ring 3, i.e. the one with the highest column density. 

Knowing the dust distances, with the assumed  GRB X-ray fluence and the derived column densities, we can compute the lightcurves of the six rings and compare them with that of the burst afterglow.  This is shown in Figure~\ref{dust_lc}, where we plot the lightcurves of the single rings and of their sum for the time interval from  1000 s to 10$^6$ s after the GRB. The afterglow light curve in the 0.3--10 keV range,  based on  the \xmm\ and {\it Swift/XRT} data\footnote{Obtained by the online tool described in \citealt{evans09}. See http://www.swift.ac.uk/user\_objects/}, is  well described by a powerlaw decay. 
Initially, the afterglow  is much brighter than the rings, but after $\sim100$ ks the total emission of the rings becomes dominant. 
This clearly indicates that the dust-scattered radiation can be significantly detected even when the  GRB afterglow has faded at lower fluxes. However, we note that we computed these lightcurves assuming a  homogeneous distribution of the dust in the plane perpendicular to the line of sight. If the dust clouds were not homogenous, the lightcurves would differ from those shown in Figure~\ref{dust_lc}.
It is also interesting to note that, for small redshifts ($z<1$), the same light curve would have been obtained if the dust clouds, with  these relative distances and column densities, were located in the host galaxy. However, in such a case it would have been impossible to resolve spatially the rings and the whole emission would have been attributed to the afterglow.
Indeed, it has been proposed that dust-scattering effects in the GRB host galaxies can, in some cases, be responsible for features seen in the light curves of the afterglows \citep[e.g.][]{shao07}. In fact, there exists a sub-class of GRBs, characterized by plateaus in the early phases after the GRB pulse, a softening of their X-ray flux on longer time-scales and long-lasting emission, where dust-scattering processes has been shown to be important \citep[e.g.][]{zhao14,evans14,wang16}.

To compare the properties we derived for the dust in the direction of \src\  with the Galactic distribution of   molecular clouds, we  used the  H I and H$_2$ surveys of \citet{dame01} and \citet{kalberla05} (Figure~\ref{himap}). The H$_2$ profiles show that most of the material is located within less than 3 kpc, with a peak at $\sim$2 kpc. On the other hand, the H I profile presents a significant peak at $\sim$0.5--1 kpc and indicates also the existence of other material in the range 2--6 kpc. The recent paper of \citet{rice16} investigated all the molecular clouds with masses $>10^4$ M$_{\odot}$, but none were found along the \src\ direction.

The first peak in the H I profile can be associated to the four closest clouds identified in our work, which contain most of the dust along the line of sight. The H I profile shows also an extended region with lower density from about  2 to 6 kpc, which can be tentatively associated to the dust-layer at $\sim$5.1  kpc (ring 6). This layer is also the one with the lowest column density ($\sim7\times10^{20}$ cm$^{-2}$) and the corresponding ring has the hardest spectrum. It therefore contains a small amount of dust grains and we could find evidence of it only because its radial expansion was still small during the \xmm\ observation (which, in turns, implies a large cross-section). In particular, the broad peak in the pseudo-distance distribution indicates that this dust-layer may be significantly extended. We can locate this dust inside the Perseus spiral arm of the Galaxy and we speculate that this may be a large region of diffuse dust rather than a single cloud.
Although definitely fainter and very poorly constrained, we cannot exclude that also the tentative rings detected at $\sim$2.5 kpc and $\sim$3 kpc might be associated to the same large Galactic structure.
Instead, from the H$_2$ profile, the bulk of the density is located closer  than 700 pc or between $\sim$1 and 3 kpc. The dust-layers of rings 1 to 4 could be associated to the first bunch of material, while the dust-layer detected at $\sim$1.5 kpc to the second one. {In addition, we note that the Galactic micro-quasar V404 Cyg (at RA, Dec = (315$\fdg$30, 42$\fdg$22)) is about 11{\textdegree} far from \src, and the line of sight towards both sources intersects the Orion Spur. In 2015, after a bright outburst, five expanding rings were observed around V404 Cyg \citep{beardmore16,heinz16} and the corresponding dust-layers located at distances of $\sim1200, 1500, 1600, 2050, 2100$ pc. We can tentatively associate the dust-layers at $\sim$1500 and 1600 pc found for V404 Cyg with the one responsible for ring 5 in this work, but, considering the elongated structure of the Orion spur and that 11 deg at distances between 500 and 2000 pc correspond to a separation ranging from $\sim$100 to $\sim$400 pc, we do not expect a close correspondence between the clouds detected along these relatively distant lines of sight.}

Finally, the Pan-STARSS 1 survey provides a three-dimensional picture of the interstellar reddening ($E(B-V)$), over 3/4 of the sky out to a distance of several kpc  \citep{green15}. We investigated the extinction in a region of 500$'$ $\times$500$'$ around the direction of \src,  by examining the maps for various ranges of distances  centered at the values derived for the dust layers (see figure~\ref{pannstar}). The maps show that the  distribution of the Galactic medium in this direction is complex, with filaments and asymmetries. High levels of extinction are seen at distances of 500 pc, 500--630 pc, 630--790 pc and 790--1000 pc, while at larger distances the extinction is weaker. In particular, we note that the largest extinction is in the range 630--790 pc, i.e. around the distance of the dust layer of ring  3, which is the one with the highest column density. 
This demonstrates that the two approaches provide similar results, further corroborating the  analysis presented here. However, we highlight that our approach in this specific case can be more powerful. In fact, the Pan-STARSS 1 extinction map is strongly dependent on the existence of known stars at the different distances. Instead we just need to know the expansion law of the dust-scattering rings, allowing us to map the Galaxy with an extremely high resolution.
Finally, we note that in this work we assumed isotropy for the dust-layers. This does not change the derived distances but affects the values of the column densities of the dust layers. Although beyond the scope of this paper, we found possible hints of azimuthal asymmetry in the rings, due to more dust in particular towards the direction of the Galactic plane (right-hand side in figure~\ref{pnimage}). This would be, again, consistent with the extinction maps of the Pan-STARSS 1 survey. However we note that the constraints on the azimuthal asymmetry are very weak and no robust claim can be {made}. 

\section*{Acknowledgements}
This work has been partially supported through financial contribution from  PRIN INAF 2014.
The results are  based on observations obtained with XMM-Newton, an ESA science mission with instruments and contributions directly funded by ESA Member States and NASA, and on data obtained from the HEASARC archive. PE acknowledges funding in the framework of the NWO Vidi award A.2320.0076. EC acknowledges support from a VIDI grant from the Netherlands Organisation for Scientific Research (NWO).
We thank Daniele Vigan\`o who developed part of the software used in this work.

\addcontentsline{toc}{section}{Bibliography}
\bibliographystyle{mn2e}
\bibliography{biblio}

\bsp
\label{lastpage}
\end{document}